\documentclass[final,5p,times,twocolumn]{elsarticle}

\usepackage{graphics}
\usepackage{epsfig}
\usepackage{epstopdf}

\usepackage{amssymb}

\journal{Physics Letters A}

\begin{document}

\begin{frontmatter}



\title{Fermiology of 122 family of Fe-based superconductors: An {\it ab initio} study}


\author{Smritijit Sen and Haranath Ghosh}

\address{Indus Synchrotrons Utilization Division, Raja Ramanna Centre for Advanced Technology, Indore 452 013, India.}

\begin{abstract}
Fermiology of various 122 systems are studied through first 
principles simulation. Electron 
doping causes expansion of electron and shrinkage of hole Fermi pockets. 
Isovalent Ru substitution (upto 35\%) makes no visible modification in 
the electron and hole like FSs providing no clue regarding the nature of charge carrier 
doping. However, in case of 32\% P doping 
there are considerable changes in the 
hole Fermi surfaces (FSs). From our calculations, it is very clear that two 
dimensionality of FSs may favour electron pair 
scattering between quasi-nested FSs which has important bearings in various orders 
(magnetic, orbital, superconducting) present in Fe-based superconductors.
\end{abstract}


\end{frontmatter}


\section{Introduction}
\label{}
\par Discovery of iron based superconductors (SCs) unwrapped a new era of superconductivity 
which indulges researchers to resolve the unsettled mysteries of high temperature 
superconductivity that remain so far unexplained within the framework of 
phonon mediated conventional BCS theory. 
Close proximity of superconductivity to structural and magnetic transitions, glue for the 
superconducting pairing, etc. are concomitant to electronic structure near Fermi energy. 
The Fermi surface (FS) is simply the surface in momentum space where, 
all the fermionic states with (crystal) 
momentum $|k|<|k_F|$ are occupied, and all the higher momentum states are empty. 
Fermi surface of Fe-based SCs mainly comprises of Fe d-orbital. Fe-based SCs have a very unusual 
Fermiology, which is very sensitive to doping, pressure and temperature. Angle-resolved 
photo-emission spectroscopy (ARPES) is one of the proficient 
experimental techniques by which FSs can be mapped. With the purpose of elucidating the 
relevance of Fermiology to the superconducting pairing mechanism 
from an experimental standpoint, several ARPES measurements have 
been performed on iron-based superconductors \cite{Lu,Malaeb,Evtushinsky}. 
The role of FS in the development of understandings in superconductivity 
cannot be overemphasized \cite{Stewart,Wang,Kontani,IImazin,Hirschfeld,Chubukov}. Electronic states
of Fe-based superconductors and superconducting (SC) pairing mechanism are very much
different from that of the high-$T_c$ cuprates \cite{Wang,Mazin}. 
FS measurement is one of the probing ingredients 
to understand the nature of pairing of electrons in Fe-based SCs. Shape of the Fermi surface 
is very crucial as it 
determines the degree of nesting in Fe based SCs which in turn give rise to magnetic 
and orbital ordering \cite{ghosh}. Unlike spin fluctuation mediated 
superconductivity in cuprates and heavy fermions, orbital fluctuation has also 
been proposed as one of the 
possible pairing mechanism of superconductivity in these 
materials \cite{Kontani,IImazin,Kuroki}. Orbital fluctuations 
are supposed to be resulting from the inter-band nesting between hole and electron 
like Fermi surfaces (FSs) and inter-orbital quadrupole interaction 
which is related to electron phonon interaction \cite{book}. Angle-resolved photo emission spectroscopy 
(ARPES) measurements, have been conducted on iron-based superconductors, 
especially for 122 family to reveal the Fermiology of Fe-based superconductors because
of availability of large numbers of high quality single crystals in this series. 
Theoretically calculated FS of 122 family is also available in 
literature \cite{Mazin,Nekrasov,Graser,Wang2,Pan}.

 The 122 family like BaFe$_2$As$_2$, CaFe$_2$As$_2$, SrFe$_2$As$_2$ {\it etc.} comprises
the heart of Fe-based superconductors where best quality single crystals are available.
All these materials exhibit spin density wave (SDW) order below a transition temperature 
and incidentally at the same 
temperature the structural transition from tetragonal to orthorhombic 
(low temperature) phase occurs. Although these parent compounds of 122 family are not 
superconducting but 
superconductivity can emerge either by applying external pressure or by chemical 
doping. In fact doping can be introduced at any of the sites. When Ba is replaced by K 
and Fe is replaced by Co, it introduces hole \cite{Rotter} and electron 
doping \cite{Sefat} in the system respectively. 
The introduction of extra hole or electron shifts the chemical potential in 
Ba$_{1-x}$K$_x$Fe$_2$As$_2$ 
\cite{Rotter} and BaFe$_{2-x}$Co$_x$As$_2$ \cite{Sefat} in such a way that the size of 
the electron and hole like 
FSs evolve oppositely which diminishes the nesting between them, resulting suppression of 
spin density wave (SDW), orbital density wave (ODW) orders and SC emerges. What change in 
FS is then expected for 
isovalent substitutions? (like Ru substitution in Fe site or P substitution in As site) 
The observation of suppression of SDW order with isovalent substitution in 
BaFe$_2$(As$_{1-x}$P$_x)_2$ and BaFe$_{2-x}$Ru$_x$As$_2$ systems is therefore, still 
under debate \cite{dhakadop,Brouet,Thaler,Ren,Ye}. We show that there are no significant 
changes either in electron or 
hole like FSs upon substantial Ru substitution and this observation is consistent 
with the observations of Dhaka {\it et al.,}\cite{dhakadop}. 
But at about 50\% doping there is 
significant change in the hole like FSs which is also consistent with the experimental 
findings by N. Xu {\it et al.,}\cite{Xu}. We study on the nature of FSs when 
doped in any of the three sites mentioned above through first principles simulations. 
From our calculations we explicitly show that in case of hole doping the charge carriers 
go to hole FSs resulting expansion of hole FSs and shrinking of electron FSs and a 
reverse situation occurs in case of electron doping. In case of isovalent P substitution in 
place of As, causes substantial z-direction dispersion in the hole bands, making shape of 
some of the hole FSs more like three dimensional which is consistent with experiments 
\cite{Kasahara,Yoshida} and is believed to be responsible for nodal superconductivity. 
On the other hand, 
two dimensional FSs which are more favourable to nesting are believed to be 
responsible for high $T_c$ superconductivity in Fe-based materials \cite{Sunagawa}. 
Theoretically computed FSs of various doped “122” system are presented, 
similarities and dissimilarities are compared among them and with 
available experimental data.
\section{First principles calculation}
  In our earlier work \cite{Sharma} a detailed structural evolution 
 in Ru substituted BaFe$_2$As$_2$ pnictide 
 superconductor as a function of Ru composition and temperature was carried out through 
 combined experiment-theory study. We showed a correspondence between the anomalies 
 in the structural parameter, particularly the displacement of the As atom and the 
 anomalous change in resistivity across the spin density wave transition. 
 First principles calculations to check the veracity of the structural 
 evolution were successfully accomplished. This is significant because, one of the open 
 problems in Fe-based materials is a difference between theoretical density functional theory (DFT) based
 simulations and its matching with experimental values of $z_{As}$. Number of authors 
 have already experimentally reported about the fact that the position 
 of As is more closer to the Fe-plane than that observed (by more than 0.1 A) 
 and cannot be reproduced accurately from DFT 
 calculations \cite{IImazin,Sharma,Sen,DJSingh,Mazin2,Zhang,Yin}. On the other hand, various 
 physical properties including magnetic and superconducting transitions 
 (specially electronic structure) depend crucially on $z_{As}$ \cite{Sen}. 
 The simulated data also explained the origin of two Fe-Fe distances. 
 
  In a separate earlier work, we have performed a detailed electronic structure (band structure, 
 density of states, Fermi surface structure) calculations using the experimental 
 lattice and structural parameters $a$(T, x), $b$(T, x), $c$(T, x), $z_{As}$ (T, x) as 
 input parameters in our DFT simulations that overcome above mentioned problems \cite{Sen}.  For clarity we 
 explain our results of \cite{Sen} briefly here. We showed that the density of states at the Fermi 
 level for Fe-d (up as well as down spin) increases with temperature in the 
 same fashion as that of the $z_{As}$ {\it i.e.}, due to modifications in the electronic 
 density of states with temperature, thermal behaviour of $z_{As}$ is dictated. The difference 
 in up and down spin density of states at the Fermi level (that provides net magnetization) 
 as a function of temperature shows onset of AFM like SDW state at the same temperature 
 as that of the structural transition. Through detailed band structure calculations we 
 showed that the $d_{xz}$, $d_{yz}$ bands of the Fe-d orbital become non-degenerate 
 (leading to orbital ordering), both at the $\Gamma$ and X points, at the structural 
 transition temperature (above which they are degenerate). This difference in 
 energies (of the $d_{xz}$, $d_{yz}$ bands) at various temperatures follows exactly the 
 same temperature dependence of the orthorhombicity parameter 
 determined experimentally. So far available first principles simulations on FS 
 of Fe-based materials are based on DFT based geometrically optimized zero 
 temperature one which do not correspond to experimental situations at room
 temperature. In this work, therefore, we present our simulations on FS
 of parent 122 and various doped 122 systems using the experimental 
 lattice and structural parameters $a$(T, x), $b$(T, x), $c$(T, x), $z_{As}$(T, x) 
 at room temperature as fixed input parameters \cite{Sefat,Kasahara,Sharma,Mrotter}.

 Our first-principles {\it ab-initio} simulations of Fermi surfaces are performed using CASTEP 
 module of
 Material studio 7.0  \cite{CASTEP}, which exploits the plane-wave pseudo-potential
 method based on DFT. In our simulations the electronic exchange
 correlation is treated under the generalized gradient approximation (GGA) using
 Perdew-Burke-Enzerhof (PBE) functional \cite{PBE}. Geometry optimization has been 
 carried out for two parent compounds of 122 system 
 in which we fix $z_{As}$. Tackling small fraction of doping (Co/Ru/K/P)
 in place of Fe/Ba/As is implemented by considering virtual crystal approximation (VCA) based
 on the Mixture Atom Editor of CASTEP in Material Studio 7.0. Non-spin polarized and spin polarized
 single point energy calculations are performed for tetragonal phase with space group
 symmetry I4/mmm (No.139) using ultrasoft pseudopotentials and plane wave basis set with energy
 cut off 500 eV and self-consistent field (SCF) tolerance as $10^{-6}$ eV/atom. Brillouin
 zone is sampled in the k space within Monkhorst-Pack scheme and grid size
 for SCF calculation is $25\times25\times33$. All these calculations have been carried out
 using primitive cell. In spin polarized calculations, spin state of the two Fe atoms have
 been fixed 
 in the opposite direction so that it gives anti ferromagnetic like ordering.
 \begin{figure}[ht]
 \includegraphics[width=8cm]{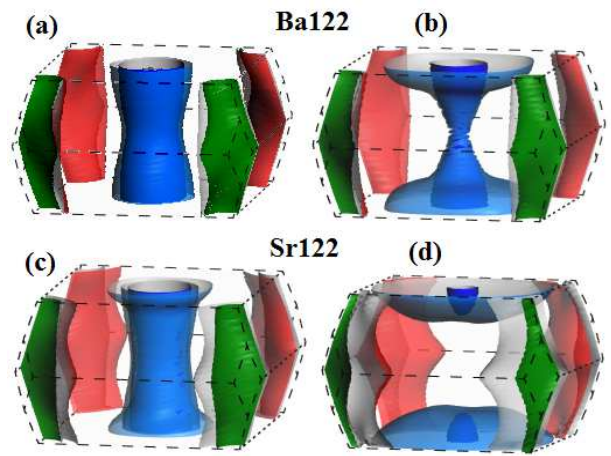}
 \caption{Calculated FSs of undoped 122 Fe-based SCs for (a) BaFe$_2$As$_2$ (b) 
optimized structure of BaFe$_2$As$_2$ (c) SrFe$_2$As$_2$ and (d) optimized 
  structure of SrFe$_2$As$_2$. In each figure FSs at the center are hole like FSs, shaded 
 blue and FSs at the corners are electron like FSs, shaded red and green.}
 \label{FS1}
 \end{figure}
\section{Results and discussions}

\begin{figure}[ht]
    \includegraphics[width=8cm]{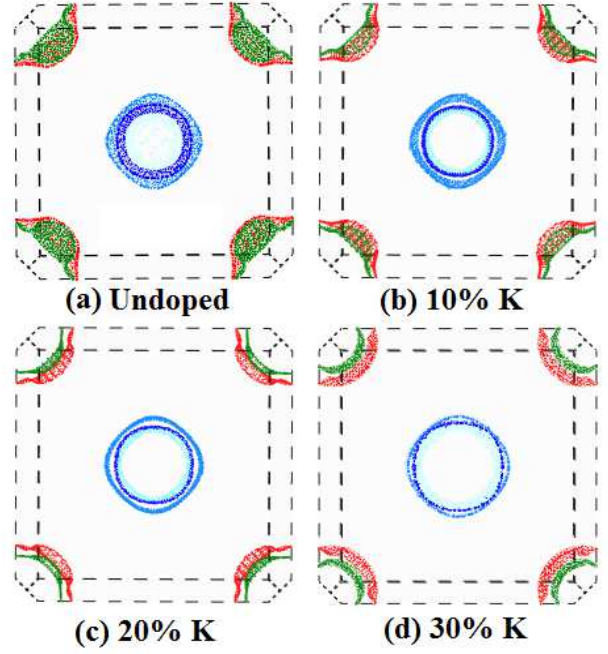}
    \caption{Top view of the calculated FSs of K doped BaFe$_2$ As$_2$ system 
    for various K doping concentration (in k$_x$-k$_y$ plane) along k$_z$ direction.}
    \label{FS3}
\end{figure}

\begin{figure}[ht]
  \includegraphics[width=8cm]{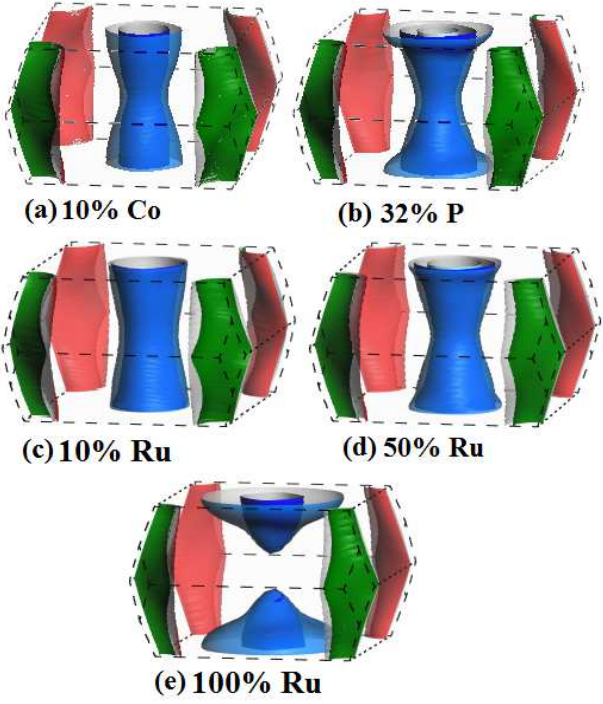}
  \caption{Fermi Surface of various doped 122 system: (a) 10\% Co doped (b) 32\% P 
  doped (c) 10\% Ru doped(d) 50\% Ru doped BaFe$_2$ As$_2$. (e) Calculated FS of BaRu$_2$As$_2$}
  \label{FS4}
  \end{figure}

\begin{figure}[ht]
 \includegraphics[width=8cm]{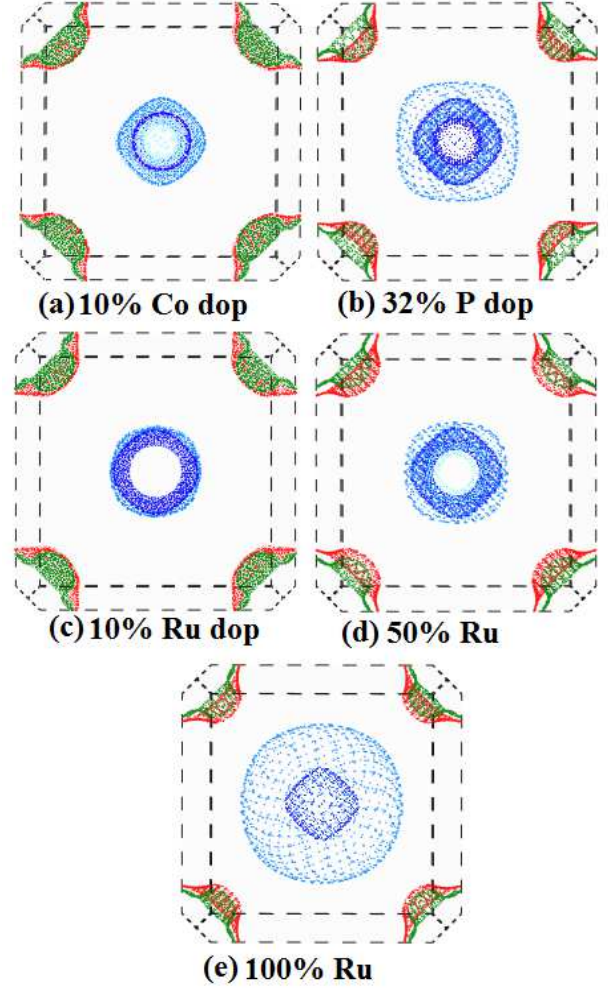}
 \caption{Top view of the FSs of various doped “122” system (in k$_x$-k$_y$ plane) along k$_z$ direction.}
 \label{FS5}
 \end{figure}

We present simulated FS structures of 122 family of Fe-based superconductors, namely, 
BaFe$_2$As$_2$, SrFe$_2$As$_2$, Ba$_{1-x}$K$_x$Fe$_2$As$_2$, BaFe$_{2-x}$Ru$_x$As$_2$, BaFe$_{2-x}$Co$_x$As$_2$, 
BaFe$_2$As$_{2-x}$P$_x$.
FSs have been calculated for 122 parent compounds BaFe$_2$As$_2$ 
and SrFe$_2$As$_2$ for unoptimized as well as optimized structures. Figure \ref{FS1} 
depicts the simulated FSs of BaFe$_2$As$_2$ and SrFe$_2$As$_2$ where three 
hole like FSs appear at the centre of the brillouin zone and two electron like FSs appear 
at the four corners of the brillouin zone. Calculated FS topology for both 
parent compounds are very similar. There are experimental evidence of SDW ordering in both the parent
compounds at low temperature. This SDW ordering is the result
of nesting between electron and hole like FSs \cite{ghosh}. 
Two dimensional nature of the FSs 
in both the systems enhances nesting and results SDW ordering. 
 \begin{figure}[ht]
  \includegraphics[width=8cm]{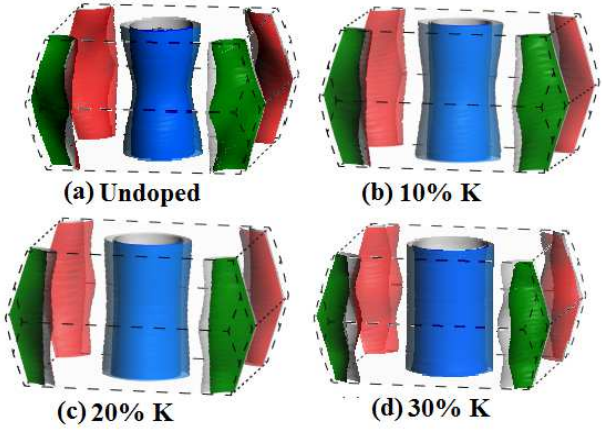}
  \caption{Simulated FSs of K doped BaFe$_2$As$_2$ system for various K doping concentration}
  \label{FS2}
  \end{figure}
 Whereas FSs generated from
 optimized structure is more like three dimensional which works against nesting and reduces the possibility 
 of SDW ordering. So FSs calculated using experimental lattice parameters provide more realistic FSs 
 that resembles with the experimentally observed one \cite{Ding, Xie}. 
 Even non-magnetic calculations (non spin polarized) do not make topologically appreciably different FSs. 
 We have also studied various doped “122” systems using 
 VCA approach. Fe d-orbital mainly constitutes the FSs of Fe based SCs. We found that 
 at larger doping in the Fe site, VCA fails and deviates from the actual FS topology. So
 for some of the calculations we adopt super cell method for example 50 \% Ru doping.
 Figure \ref{FS2} and Figure \ref{FS3} illustrate simulated FSs of K-doped (hole doping) 
 systems at various K doping concentrations. It is very clear from 
 Figure \ref{FS2} and Figure \ref{FS3} that 
 K doping exerts extra holes to the system which expand the hole like FSs at the centre 
 whereas electron like FSs shrink at the corners. 
 Exactly the reverse scenario occurs for Co doping (see Figure \ref{FS4}a and \ref{FS5}a).
 A more closer look on Figure \ref{FS2} and Figure \ref{FS3} reveals that there are significant changes in 
 electron and hole like FSs upon K doping. With increasing K concentration the $k_z$ dispersion 
 of one of the hole FS (outer one) becomes more and more weaker whereas around 30 \% doping 
 all the hole FSs take the shape of perfect cylinder. Electron FSs also evolve significantly 
 with doping and shows weaker $k_z$ dispersion with increasing doping concentration. 
 Thus a transition in FS structure from quasi 3 dimensional to quasi 2 dimensional structure with doping is observed.
 As the hole and electron FSs evolve oppositely with K doping, the overall nesting 
condition degrades due to 
 size mismatch of the FSs resulting in suppression of SDW order but at the 
 same time two dimensionality of the FSs in these systems favours very large
density of states at the Fermi level, enhancing the electron pairing
possibilities. 
 So a competing order of superconductivity and magnetism  are likely to co-exist \cite{Li}. 
 Using two band model, it was shown that apart from 
 inter band paring, superconductivity can also arises from intra band paring and combined intra-inter 
 band pairing results in higher $T_c$ compared to only inter band scenario \cite{JALCOM}. 
 The effect of intra band scattering also depends on the topology of FS. These facts probably
 explain the reason of significantly higher $T_c$ in K-doped systems compared to other 
 122 systems. In case of Co doping with increasing doping concentration the 
 electron FSs expand and hole FSs shrink whereas the $k_z$ dispersion is stronger. It is presented in
 figure \ref{FS4}a and figure \ref{FS5}a which is consistent with 
 earlier theoretical and experimental results \cite{Mazin}. This 
 suppresses magnetic ordering and emerges superconductivity into these systems. On the other hand, 
substantial P doping in place of As, (like 32\%) modifies hole FSs considerably as shown in 
 figure \ref{FS4}b and figure \ref{FS5}b. In contrast to K doping, 
 P-doping in place of As, a clear dimensional cross-over from two to three dimensions is 
observed in the FS structure consistent with reference \cite{Jiang}. 
 However, no significant changes in both the electron and hole FSs occur upon Ru substitution 
 up to 35\%, as was also found in experiments \cite{dhakadop}. At about 50\% Ru doping 
 there is significant modifications in the hole like FSs; more precisely, a dimensional crossover 
 from two to three dimension occurs which is also consistent with recent experiment \cite{Xu}. 
 But calculated FS of BaRu$_2$As$_2$ indicates complete loss of 
 two dimensionality of hole FSs which in turn causes large degradation of nesting. 
 This explains the absence of magnetic order (SDW) in BaRu$_2$As$_2$ \cite{Zhang}. 
 In figure \ref{FS4}c, \ref{FS4}d and \ref{FS4}f,
 evolution of FSs with Ru doping has been depicted. In figure \ref{FS5}a ``top view" of all 
 the FSs of figure \ref{FS4} are represented to illustrate 
 the changes in the sizes of the Fermi pockets. 
 A closer look at all the 
 Fermi pockets (Figure \ref{FS5}) reveal that there are two hole 
 like FSs in most of the cases, one symbolized by light blue ring (outer) 
 and other one by deep blue ring (inner). The inner and outer radii of any of 
 the ``Fermi rings" denote the radii of the FSs around $\Gamma$ and Z points respectively 
 ($k_z$ dispersion). These are experimentally measurable through ARPES studies 
 and are found to be consistent with some of the observed momentum distribution 
 curves \cite{dhakadop}.
 Momentum distribution curves in general, are useful to estimate
both components of the electronic self energy (real and imaginary parts)
 which is a powerful aspect of ARPES study. The momentum distribution 
curves of Fe-based materials are directly related to the Fermi surface radii because,
it is a measure of electronic band structure dispersion width of a given band \---- how close
or far (thus proportional to $\Delta k$) the electronic band intersects two Fermi points.
 In case of Co, P doping, changes in the sizes of the FSs 
 and dimensionality cross-over (3D) reduce nesting and thus suppress magnetic order, e.g., 
 BaRu$_2$As$_2$. All these results 
 indicate a possible competition between superconductivity and magnetic order (SDW). 
 SDW order (or orbital order) is controlled by inter band nesting of FSs which in 
 turn is related to the dimensionality of the FSs. Emergence of superconductivity
 in these Fe-based systems seem to be originated from spin fluctuation or orbital 
 fluctuation which may be enhanced by quasi nesting of two dimensional 
 electron and hole FSs as suggested by M. Sunagawa {\it et al.,}\cite{Sunagawa}.
 Since the FSs are Fe-d orbital derived, Co-doping (which has one more d-electron than Fe) 
clearly causes disturbance and makes modification
to the electronic and hole FSs. On the other hand, P doping in place of As does not cause
change in the d-orbital occupancy of Fe directly, but it causes changes in the pnictide
height causing a z-direction dispersion. Similarly, Ru having larger atomic dimension than
Fe, when replaced, the in-plane lattice parameter $a$ increases whereas the out-of-plane lattice 
parameter $c$ decreases
after certain amount of doping, affecting z-dispersion in band structures. Both the 
above two (modifications either in the pnictide height or lattice constants as described above)
are not expected in case of K doping in place of
Ba. These are the possible origins for different nature of
 dimensional cross-over in the
FS structures with substitutions in various 122 systems.

\section{Conclusions}

Calculated FSs of the parent compounds of 122 systems are very similar in topology.
Presence of quasi two dimensional hole and electron like FSs enhances the chance of nesting. 
This is the reason why these parent compounds display magnetic 
and orbital ordering \cite{Sunagawa,Kordyuk}. 
A dimensional cross-over, in theoretically computed FS topologies of 
various Fe-based materials of ``122'' family are presented with various kinds of doping. 
Our results are consistent 
with experimental observations and its possible significance to magnetism and SC are 
presented.  In case of electron and hole doping, sizes of the electron and hole 
Fermi pockets evolve oppositely with increasing doping concentration. Isovalent 
Ru substitution up to certain doping concentration makes no visible modifications in the electron 
and hole like FSs but in case of 32\% P doping there are substantial changes in 
the hole FSs. 100\% Ru substitution modifies hole FSs remarkably. From all these 
calculated FSs it is very clear that dimensionality of FS 
(linked with FS nesting) plays an important role 
in 122 Fe-based SCs. 

\section{Acknowledgement}
One of us (SS) acknowledges the HBNI, RRCAT for financial support and encouragements. We 
are grateful to Dr. G. S. Lodha and Dr. P. D. Gupta for their encouragements and support. \\




\bibliographystyle{model1a-num-names}
\bibliography{<your-bib-database>}



\end{document}